\documentclass[twocolumn,prl,aps,showpacs,superscriptaddress]{revtex4-1}
\usepackage{graphicx,graphics}
\usepackage{amsmath}
\usepackage{amssymb}
\usepackage{epstopdf}
\usepackage{amstext}
\usepackage{amsthm}
\usepackage{amsfonts}
\usepackage{latexsym}
\usepackage{array}
\usepackage{xfrac}
\usepackage{color}
\usepackage[toc,page]{appendix}
\usepackage{subfigure}
\newcommand{\igbjd}[1]{}\newcommand{\beqa}{\begin{eqnarray}}
\newcommand{\eeqa}{\end{eqnarray}}
\newcommand{\beq}{\begin{equation}}
\newcommand{\eeq}{\end{equation}}
\usepackage[normalem]{ulem}

\usepackage{color}

\begin{document}

\title{Quantum droplets of bosonic mixtures in a one-dimensional optical lattice}
\author{Ivan Morera}
\affiliation{Departament de F\'isica Qu\`antica i Astrof\'isica, Facultat de F\'{\i}sica, Universitat de Barcelona, E--08028 Barcelona, Spain}
\affiliation{Institut de Ci\`encies del Cosmos, Universitat de Barcelona, ICCUB, Mart\'i i Franqu\`es 1, Barcelona 08028, Spain}
\author{Grigori E. Astrakharchik}
\affiliation{Departament de F\'isica, Universitat Polit\`ecnica de Catalunya,Campus Nord B4-B5, E-08034 Barcelona, Spain}
\author{Artur Polls}
\affiliation{Departament de F\'isica Qu\`antica i Astrof\'isica, Facultat de F\'{\i}sica, Universitat de Barcelona, E--08028 Barcelona, Spain}
\affiliation{Institut de Ci\`encies del Cosmos, Universitat de Barcelona, ICCUB, Mart\'i i Franqu\`es 1, Barcelona 08028, Spain}
\author{Bruno Juli\'{a}-D\'{i}az}
\affiliation{Departament de F\'isica Qu\`antica i Astrof\'isica, Facultat de F\'{\i}sica, Universitat de Barcelona, E--08028 Barcelona, Spain}
\affiliation{Institut de Ci\`encies del Cosmos, Universitat de Barcelona, ICCUB, Mart\'i i Franqu\`es 1, Barcelona 08028, Spain}
\affiliation{ICFO-Institut de Ciencies Fotoniques, The Barcelona Institute of Science and Technology, 08860 Castelldefels (Barcelona), Spain}

\date{\today}

\begin{abstract}
We demonstrate the existence of quantum droplets in two-component one-dimensional Bose-Hubbard chains. The droplets exist for any strength of repulsive intra-species interactions provided they are balanced by comparable attractive inter-species interactions. The ground-state phase diagram is presented and the different phases are characterized by examining the density profile and off-diagonal one- and two-body correlation functions. A rich variety of phases is found, including atomic superfluid gases, atomic superfluid droplets, pair superfluid droplets, pair superfluid gases and a Mott-insulator phase. A parameter region prone to be experimentally explored is identified, where the average population per site is lower than three atoms, thus avoiding three-body losses. Finally, the bipartite entanglement of the droplets is found to have a non-trivial dependence on the number of particles.
\end{abstract}
\maketitle

Ultracold gases trapped in optical lattices provide highly controllable setups which nowadays implement versatile quantum simulators for quantum many-body problems~\cite{RevModPhys.80.885,lewenstein2012ultracold}. A prominent example is the experimental observations of a quantum zero-temperature phase transition (QPT)~\cite{sachdev2011quantum} between a weakly interacting superfluid Bose gas and a strongly interacting Mott-insulator in three~\cite{Greiner2002,PhysRevLett.92.130403} and one dimensions~\cite{Haller2010}

Recently, a whole new class of ultra-dilute droplet-like quantum liquids has been studied both theoretically~\cite{PhysRevLett.115.155302,PhysRevA.94.021602,PhysRevA.94.043618,Kartashov2019} and experimentally~\cite{PhysRevLett.116.215301,Schmitt2016,PhysRevX.6.041039,Cabrera301,PhysRevLett.120.135301,PhysRevLett.120.235301,PhysRevResearch.1.033155}. Such quantum droplets are self-bound objects and are capable of existing with no external trapping, similarly to the case of helium droplets~\cite{Barranco2006}. The crucial difference is that the ultracold atoms provide an unprecedented control over the tunability of interactions and geometry of the system. The droplets were first produced in Bose gases with dipolar interactions~\cite{PhysRevLett.116.215301,Schmitt2016,PhysRevX.6.041039,PhysRevLett.95.190406} and afterwards in binary bosonic mixtures with contact-like interactions~\cite{Cabrera301,PhysRevLett.120.135301,PhysRevLett.120.235301,PhysRevResearch.1.033155}. The observed equilibrium density can be eight orders of magnitude smaller than in liquid helium, due to an almost exact compensation between mean-field repulsion and attraction. Moreover, it was evidenced~\cite{PhysRevLett.115.155302} that the existence of the quantum droplets itself is a beyond mean-field effect as on the mean-field level the system collapses.

Arguably, the one-dimensional (1D) case is very promising. On one hand, in 1D quantum effects are enhanced, and, on the other hand, stability is increased due to the suppression of three-body losses~\cite{PhysRevLett.90.010401, PhysRevA.68.031602,PhysRevLett.92.190401, PhysRevLett.95.190406}. Contrary to the 3D case, the quantum droplets appear in the 
regime where at the mean-field level the system is on average repulsive and predicts a stable gas~\cite{PhysRevLett.117.100401}. In this case, quantum fluctuations result in an effective attraction which is able to liquefy the system for an arbitrary number of 
particles~\cite{PhysRevA.98.013631}.

In this Letter we go one step further and demonstrate the existence of quantum droplets in 1D Bose-Hubbard mixtures for both small and large interaction strengths. Computing the ground state with DMRG methods, we obtain the phase diagram of the system in the relevant parameter region where droplets can be produced. A very rich phase diagram is obtained, with atomic and pair superfluids appearing in both droplet and gas phases. This exciting scenario is found to be within imminent reach with current experimental setups. 

{\bf Two-component Bose-Hubbard model.}
We study a bosonic mixture with short-range interactions loaded into a high 1D optical lattice at zero temperature described by the Bose-Hubbard Hamiltonian~\cite{lewenstein2012ultracold}
\begin{eqnarray}
H&=&-t \, \sum_i \sum_{\alpha=A,B} \left( \hat{b}_{i,\alpha}^{\dagger}\hat{b}_{i+1,\alpha}+\text{h.c.}\right) 
\label{eq:TwoBH}\\
&+&\frac{U}{2}\sum_i \sum_{\alpha=A,B} \left(\hat{n}_{i,\alpha}\left(\hat{n}_{i,\alpha}-1\right)\right)
+U_{AB}\sum_{i}\hat{n}_{iA}\hat{n}_{iB} \,,\nonumber
\end{eqnarray}
where $\hat{b}_{i\alpha}$ ($\hat{b}_{i\alpha}^{\dagger}$) are the annihilation (creation) bosonic operators at site $i=1,\dots,L$ for species $\alpha=A,B$, respectively, and $\hat{n}_{i\alpha}$ are their corresponding number operators. We assume an equal tunneling strength, $t>0$, and repulsive intra-species interaction strength, $U>0$, for both components. For the rest of the work we set the energy scale to the tunneling strength $t$, which is kept fixed, and work with equal number of bosons of both species, $N_A=N_B\equiv N/2$. For convenience, we introduce the dimensionless ratio $r=1+U_{AB}/U$, and concentrate on the case of attractive inter-species interaction $U_{AB}<0$ but always fulfilling $|U_{AB}|<U$ ($r>0$). This choice is motivated by previous studies in the continuum geometry, where mean-field interactions were shown to compensate each other exactly for $r=0$~\cite{PhysRevLett.117.100401}. 

\begin{figure}[t]
\includegraphics[width=0.95\columnwidth]{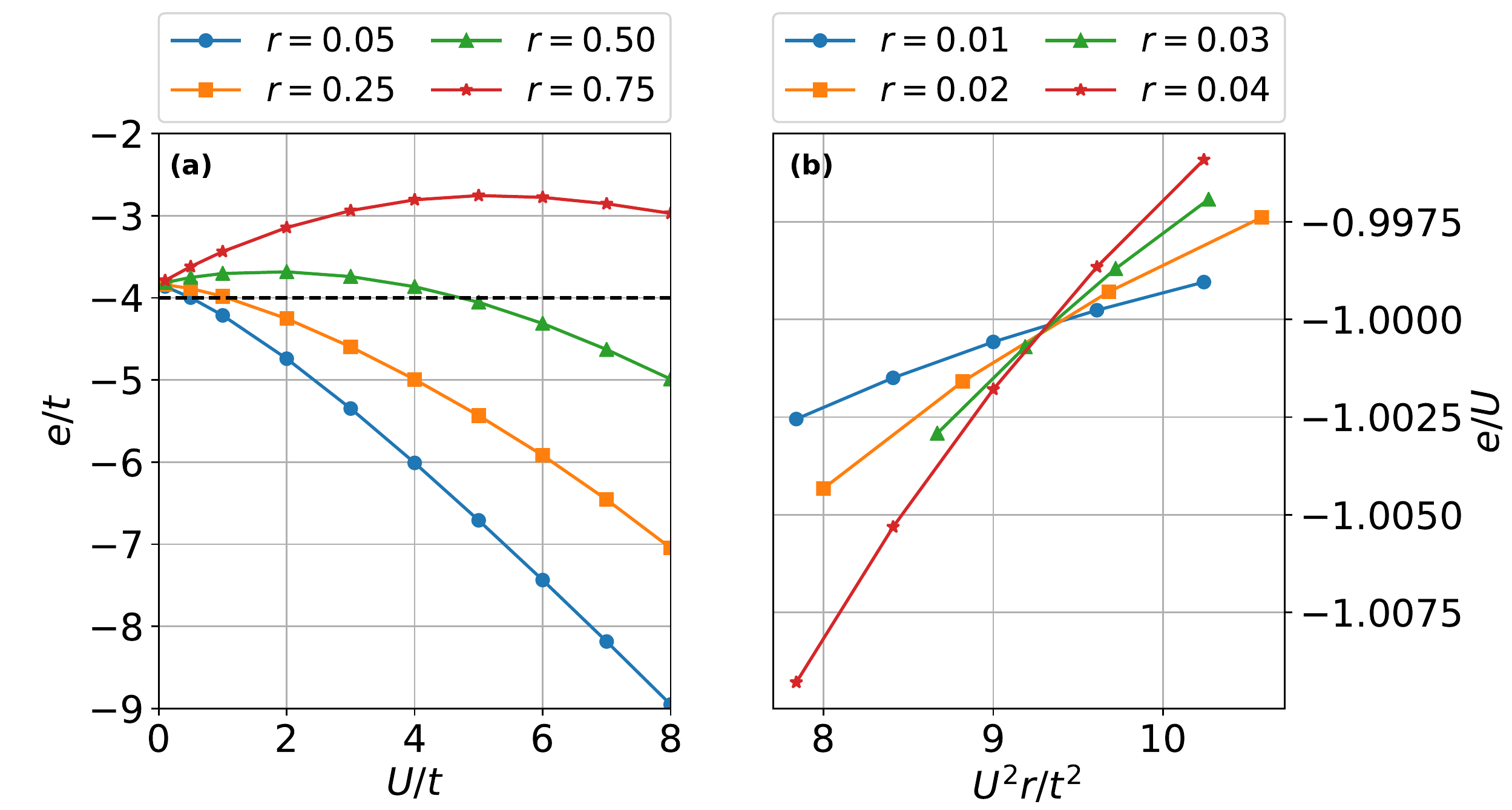} 
\caption{Energy per particle $e=2E/N$ in the system with $N=128$ as a function of the intra-species interaction strength for different values of $r$ from the weakly interacting situation (a) to the strongly interacting one (b). }
\label{Fig:EnergyN}
\end{figure}

\begin{figure}[t]
\includegraphics[width=0.95\columnwidth]{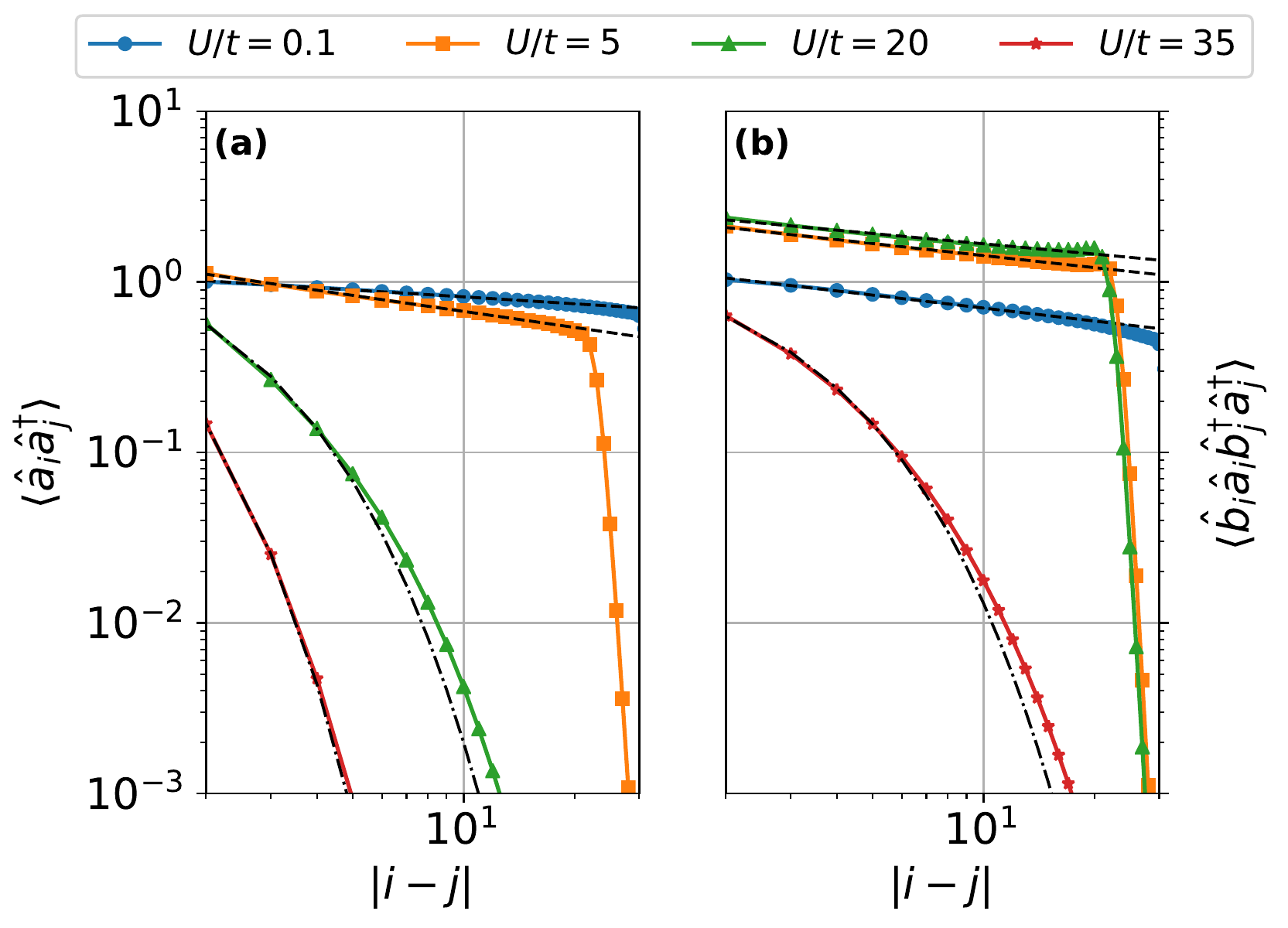} 
\caption{Correlation functions as a function of the distance $|i-j|$. The index $i=32$ is fixed at the center of the lattice of size $L=64$. Dots correspond to numerical data and dashed (dotted-dashed) lines correspond to algebraic (exponential) fits. For the droplet configurations ($U/t=5,20$), an abrupt exponential decay is visible in the correlation functions at the edges of the drop. In all cases, $r=0.01$.
}
\label{Fig:Correlators}
\end{figure}

\begin{figure}[t]
\includegraphics[width=0.95\columnwidth]{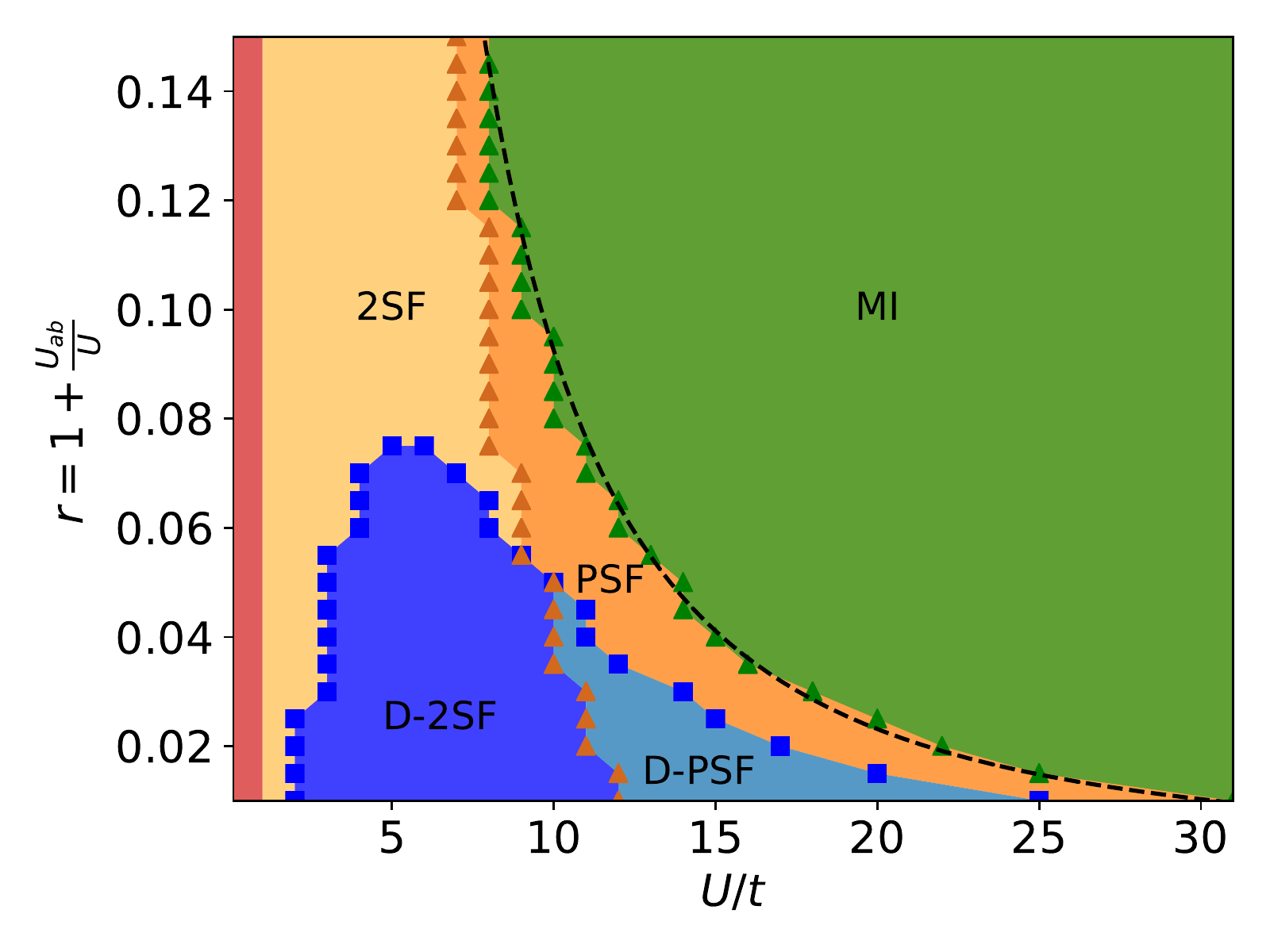} 
\caption{Phase diagram of the two-component Bose-Hubbard model Eq.~\eqref{eq:TwoBH} close to the droplet regime for $N_A=N_B=L=64$. The MI phase is characterized by the exponential loss of phase coherence. Green triangles represent the MI-PSF phase transition characterized by the appearance of quasi-long range coherence in the two-body correlators. The dashed line represents the MI-PSF transition line, $r = 9.25 \, t^2/U^2$, see the main text. The PSF-2SF transition (orange triangles) is characterized by the appearance of quasi-long range coherence in the one- and two-body correlators. Blue squares represent the regime where we detect the appearance of droplets with a size smaller than the lattice length. Droplets are identified by the exponential decay in the density at the edges.}
\label{Fig:PhaseDiagram}
\end{figure}

{\bf From Mott-insulator to pair superfluid droplets}
The Hamiltonian~\eqref{eq:TwoBH} at strong coupling $U\gg t$ supports different quantum phases, including the Mott-insulator (MI) and the pair superfluid (PSF)~\cite{PhysRevLett.92.050402,PhysRevA.80.023619}. The PSF is a state characterized by the formation of pairs of atoms (molecules) which exhibit long-range phase coherence~\cite{PhysRevLett.92.050402,PhysRevA.80.023619,PhysRevLett.92.030403,PhysRevLett.92.160402}. On the other hand, coherence 
is exponentially lost in the MI state. As predicted in Ref.~\cite{PhysRevLett.92.030403}, the transition from the MI state into the PSF takes place at a fixed value of $U^2r/t^2$ independent of the $r$, see Fig.~\ref{Fig:EnergyN}(b). We numerically extract the position of the critical point for the MI-PSF transition and obtain $(U^2r/t^2)_c \simeq 9.25$, shown with the dashed line in Fig.~\ref{Fig:PhaseDiagram}. Superfluid states are commonly characterized by having long-range phase coherence. Therefore, we test if the one- (two-) body correlation functions possess off-diagonal quasi-long range order, seen as a slow power-law decay, and interpret its presence as superfluidity of atoms (pairs). In addition, we have verified for several selected points that PSF phase possesses a finite gap $\Delta = E(N+1,N)-2E(N,N)+E(N-1,N)>0$ which instead is absent in other phases~\cite{PhysRevA.93.021605}. Thus, the PSF is characterized by: (i) absence of phase coherence in the one-body correlator, (ii) appearance of phase coherence in the two-body correlator, and (iii) finite gap associated with spin excitations. Therefore, the correlations between bosons of the same species decay exponentially with the distance $\langle \hat{a}_i \hat{a}^{\dagger}_j \rangle=\langle \hat{b}_i \hat{b}^{\dagger}_j \rangle\propto \exp{\{|i-j|/\xi\}}$, see the $U/t=20$ line in Fig.~\ref{Fig:Correlators}(a). Simultaneously, in the PSF phase the correlation function of pairs should behave as, $\langle \hat{a}_i \hat{b}_i \hat{a}^{\dagger}_j \hat{b}^{\dagger}_j \rangle\propto 1/|i-j|^{\alpha}$~\cite{PhysRevA.80.023619}, see the $U/t=20$ line in Fig.~\ref{Fig:Correlators}(b). Finally, for the MI state all correlation functions decay exponentially, see the $U/t=35$ line in Fig.~\ref{Fig:Correlators}. The change of the form of the decay from exponential in the MI phase to algebraic in the PSF one, allows one to identify the phase transition, see Fig.~\ref{Fig:PhaseDiagram}. 

\begin{figure}[t]
\includegraphics[width=0.95\columnwidth]{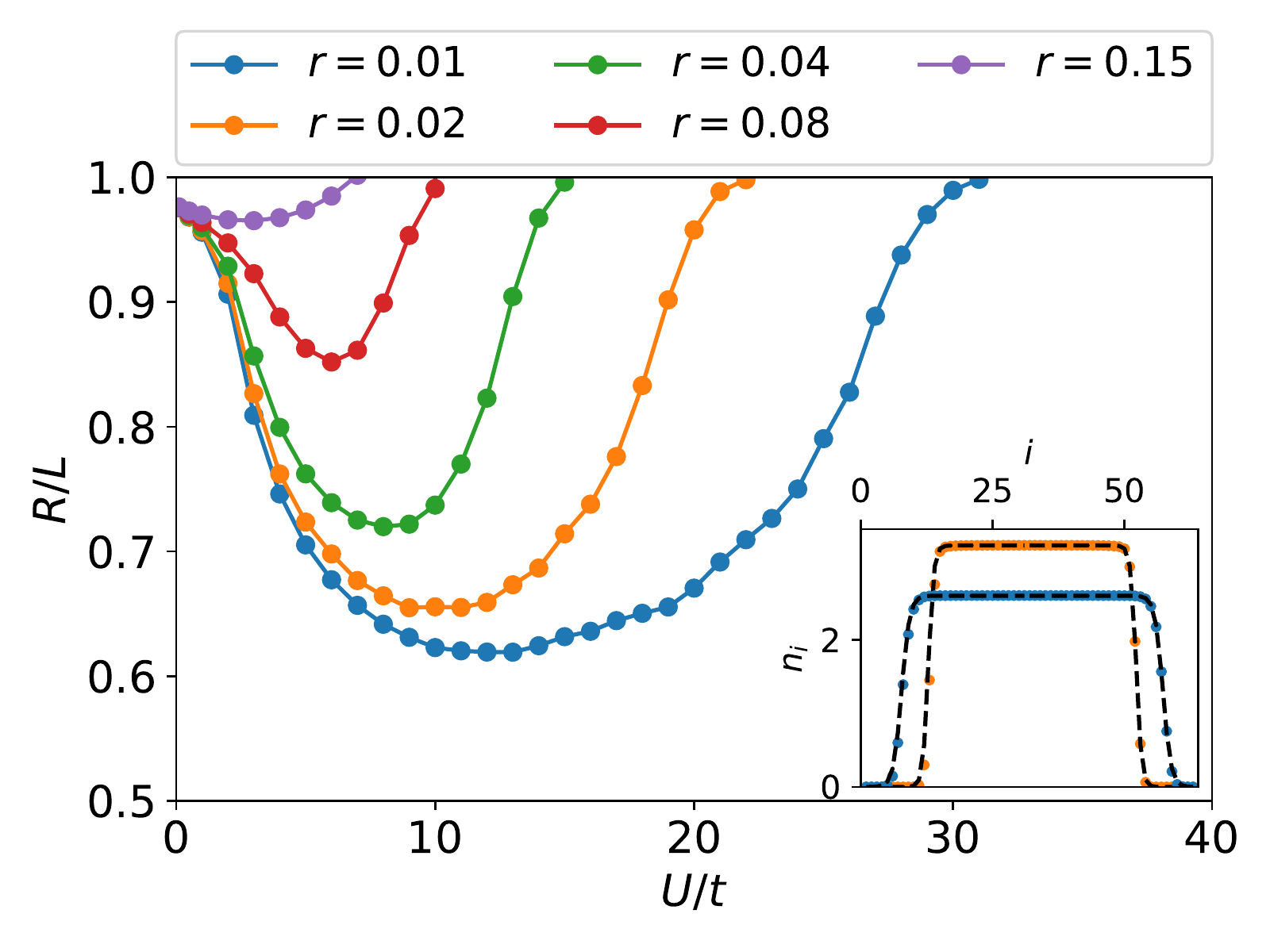} 
\caption{Typical size of the system as a function of the intra-species interaction for different ratios $r$, for $N_A=N_B=L=64$. Inset: Total density profiles of a droplet configuration (dots) for two different interaction strengths $U/t=15,25$ for $r=0.01$. The dashed lines are fits~(\ref{eq:fit}) to the density profiles. }
\label{Fig:RadiN}
\end{figure}

An intrinsic property of a droplet-like liquid is being self-bound and localized at zero pressure while a gas occupies the whole available volume. Thus, the density profile contains additional information on the phases, see inset of Fig.~\ref{Fig:RadiN} for some examples. For certain parameter values, self-bound objects are observed which do not occupy the whole available space. To characterize these objects, we fit the total density with a symmetrized Fermi function~\cite{Sprung_1997} which produces a flat-top profile of size $R$ and an exponential decay with typical length scale $a$ at the edges,  
\begin{equation}
n_i={n_{M} \sinh{(R/(2a))} \over \cosh{(R/(2a))}+\cosh{(i/a)}}  \,,
\label{eq:fit}
\end{equation}
with $n_{M}$, $R$ and $a$ being free parameters. The size of these droplets has a non-trivial dependence on the interaction $U$ for fixed $r$ as shown in Fig.~\ref{Fig:RadiN}. For any value of $r$, there is a certain value of $U/t$ above which the system is in a MI state, which extends to the full lattice. Decreasing $U/t$, the size of the droplet is found to decrease up to a critical value of $U/t$. Beyond this point, for weaker interactions the droplet size starts to increase until we reach again a size comparable to the system size in the non-interacting limit. In the following, we obtain the phase diagram and clearly identify the important differences between the weakly and strongly interacting regimes. 

{\bf The phase diagram.}
The system undergoes dramatic changes when it is brought from the strongly interacting to the weakly interacting regime. Indeed, if we decrease the interaction below the critical value discussed above, we observe that the system goes from a PSF to a two atomic superfluids (2SF) state. In the latter, the pair correlator $\langle \hat{a}_i \hat{b}_i \hat{a}^{\dagger}_j \hat{b}^{\dagger}_j \rangle$ and the two correlators $\langle \hat{a}_i \hat{a}^{\dagger}_j \rangle$ and $\langle \hat{b}_i \hat{b}^{\dagger}_j \rangle$, exhibit a power-law behavior meaning that each species separately features quasi-long-range phase coherence, see for instance the $U/t=5$ line in Fig.~\eqref{Fig:Correlators}. Thus, we can now draw two critical lines denoting the quantum phase transition from MI to PSF and from PSF to 2SF, see the phase diagram in Fig.~\ref{Fig:PhaseDiagram}. At the same time, PSF-2SF transition can occur when both phases are in their gaseous form or in their droplet configuration (termed D-PSF and D-2SF). The region of parameters $U/t$-$r$ in the phase diagram where droplet configurations (smaller than the system size) are stable is also represented in Fig.~\ref{Fig:PhaseDiagram}. Importantly, one  expects that Andreev-Bashkin drag~\cite{JETP.42.164} is maximal in the 2SF phase in the vicinity of the transition to PSF~\cite{Nespolo_2017,PhysRevLett.121.025302}. In this case, a superflow imposed on one component induces a supercurrent in the second component which is dragged without any energy dissipation.

\begin{figure}[t]
\includegraphics[width=0.95\columnwidth]{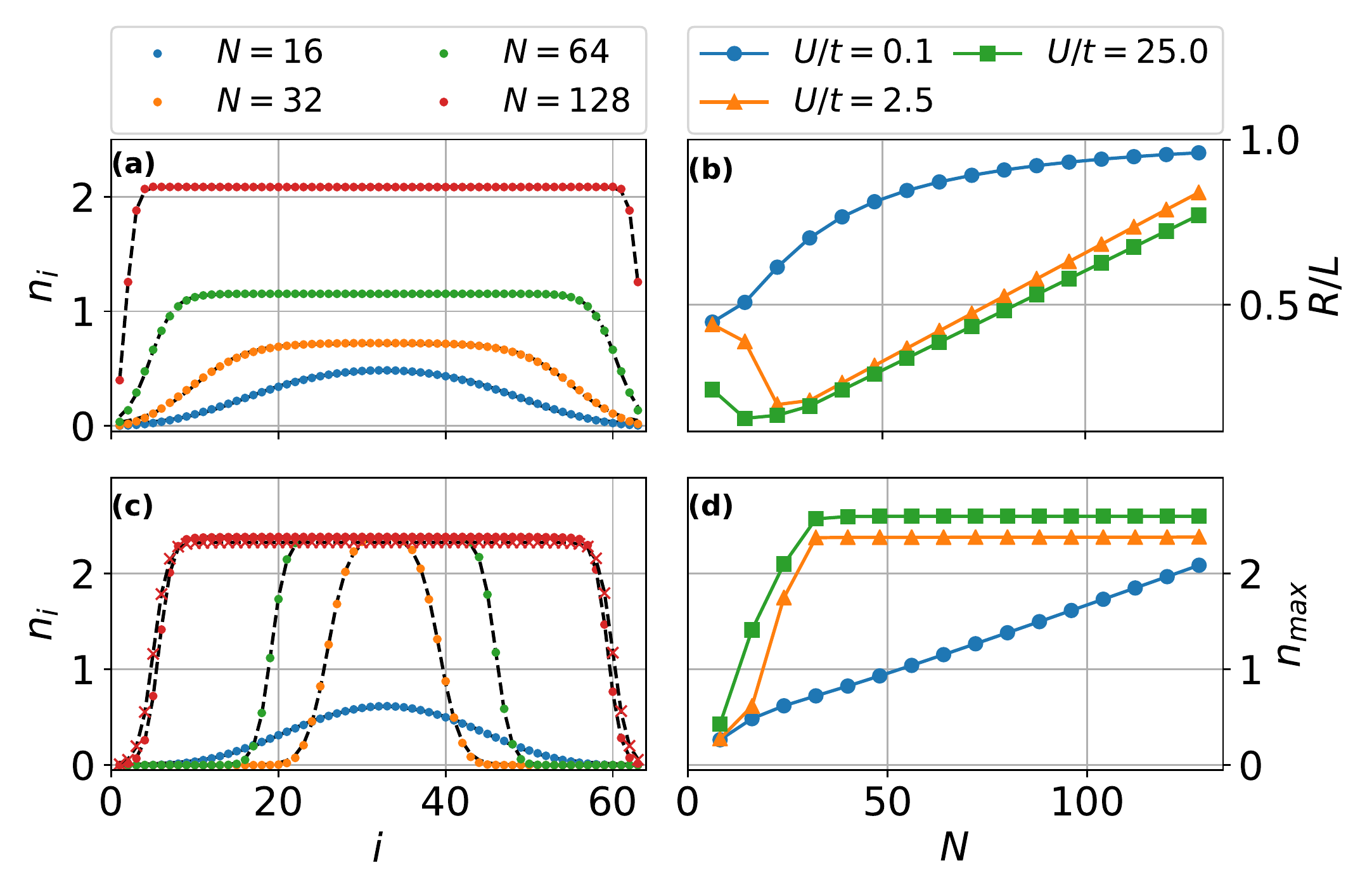} 
\caption{Panels (a) and (c): Total density profiles $n_i$ for weak interactions with $r=0.01$ and several values of $N$. Two cases are considered, a gaseous 2SF state, $U/t=0.1$, and a D-2SF, $U/t=2.5$, see dots in panels (a) and (c), respectively. An asymmetric configuration with $U_{AA}=2U_{BB}=5$ is shown with red crosses in panel (c).  Black dashed lines correspond to fits using Eq.~(\ref{eq:fit}). Panels (b) and (d): Droplet properties for a gaseous 2SF $U/t=0.1$, a D-2SF $U/t=2.5$  and a D-PSF $U/t=25$. The typical spatial size and the maximum density of the system is shown in panels (b) and (d), respectively. 
}
\label{Fig:Densities}
\end{figure}

We observe that in the 2SF phase the energy of the system decreases with increasing value of the interaction $U/t$ up to some critical value of $r$, see Fig.~\ref{Fig:EnergyN}(a). Therefore, there is a regime where the energy of the system is larger than the zero-point one $E_0=-2 N t$ where droplet configurations are not stable. This corresponds to the $U/t\lesssim 1$ region in the phase diagram, see Fig.~\ref{Fig:PhaseDiagram}.

{\bf Quantum droplet properties.}
Above we have characterized the properties of the ground state for the important case of an integer filling, $N_A=N_B=L$. Here we study how the properties change with the filling fraction in a symmetric mixture, $N_A=N_B\equiv N/2$. We consider three characteristic examples taken from different phases: gaseous 2SF, a 2SF droplet and a PSF droplet. For the gaseous configuration, atoms tend to occupy the whole available space for any number of particles $N$, see Fig.~\ref{Fig:Densities}(a). When the number of particles $N$ is augmented the maximum density $n_{\text{max}}$ and the typical extension $R/L$ always increase, see $U/t=0.1$ line in Fig.~\ref{Fig:Densities}(d). On the other hand, for the droplets a different tendency is observed allowing one to differentiate two additional regimes: small and saturated droplets. In the first regime (small number of particles) the droplets are weakly bound and have a large spatial size. An increase in the number of particles results in a stronger binding, decreasing the size and increasing the maximal density, see data for $U/t=2.5$ and $25$ in Fig.~\ref{Fig:Densities}(b,d). Then, once the density in the center of the droplet reaches the equilibrium density of the homogeneous liquid, the density saturates and a flat-top plateau is formed. In this second regime (large number of particles) the size of the droplets increases for increasing number of particles, while the maximum density remains constant and equal to the equilibrium density. In both regimes the droplets feature an exponential decay of their density profiles at their boundaries. A notable difference between 2SF and PSF droplets is seen on the small $N$ behavior: The size of PSF droplets drops abruptly as $N$ is increased, while for 2SF the behavior is smoother, as can be seen comparing the $U/t=2.5$ and $U/t=25$ lines in Fig.~\ref{Fig:Densities}(b).

\begin{figure}[t]
\includegraphics[width=0.95\columnwidth]{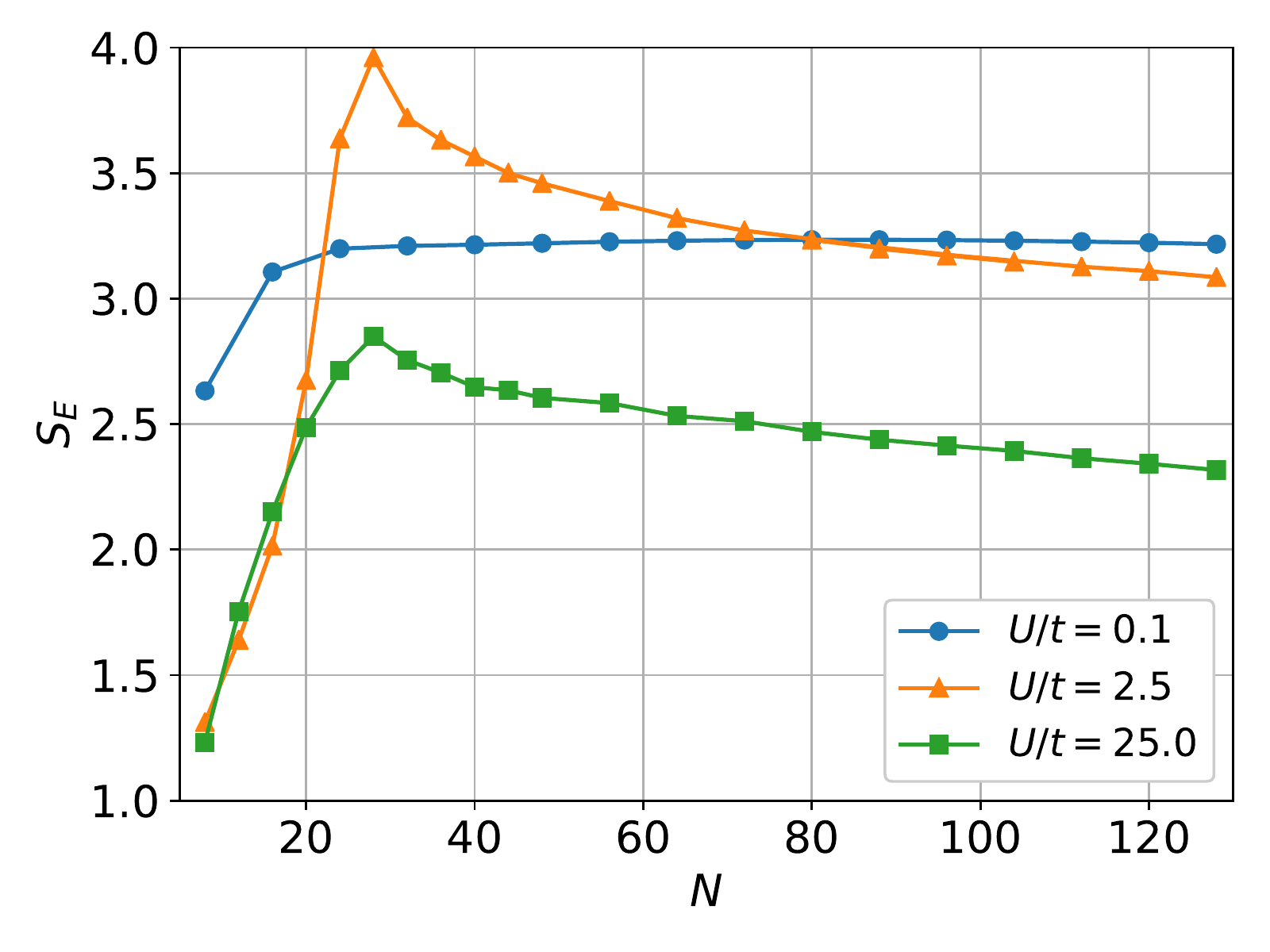} 
\caption{Entanglement entropy as a function of the total number of particles. We fixed the ratio $r=0.01$ and we choose three interaction strengths $U/t=0.1,2.5,25.0$ corresponding to a gaseous two-superfluid, a droplet two-superfluid and a droplet pair-superfluid, respectively.}
\label{Fig:EntropyN}
\end{figure}

Finally, we explore the quantum entanglement present in this system. We consider equal bipartitions of the state and explore the $\textit{left}$-$\textit{right}$ entanglement. In Fig.~\ref{Fig:EntropyN} we show the entanglement entropy corresponding to these bipartitions as a function of the total number of particles $N$ and for three characteristic values of the interaction. For the gaseous state we observe that the entanglement entropy saturates as the number of particles is increased. Instead, for the droplet configurations the regimes of saturated and non-saturated droplets can be distinguished. For small number of particles the entanglement entropy rapidly increases with the number of particles. Then it reaches a maximum at the same point where the droplets show minimum size, shown in Fig.~\ref{Fig:Densities}. For larger number of particles the entanglement entropy decreases. Therefore, we can conclude that the droplets with minimum size for a given number of particles are the ones showing larger quantum effects. It is interesting to note that this is exactly the regime of the strongest correlations where the maximum frequency of the breathing mode in absence of the lattice is reached~\cite{PhysRevA.98.013631}.

{\bf Notes on a possible experimental implementation.}
One of the main complications of the experimental observation of quantum droplets is their very short life-times due to three-body losses~\cite{PhysRevLett.120.235301,PhysRevResearch.1.033155}. This effect can be suppressed by reducing the density of the quantum droplet~\cite{PhysRevResearch.1.033155}. In our system we have found quantum droplets with densities below three for which three-body losses are greatly suppressed. Another important aspect to take into account for a possible experimental implementation is the asymmetry between the two bosonic species~\cite{Cabrera301,PhysRevResearch.1.033155}. To this aim, we have performed numerical simulations introducing an asymmetry between the intra-species interaction strenghts $U_{AA}/t=2 U_{BB}/t$. In this case, the droplet remains stable and the density remains below three, see Fig.~\ref{Fig:Densities}. All this together makes the system under study very suitable for the possible experimental observation of long-lived quantum droplets.

In conclusion, we have demonstrated the existence of quantum droplets in two-component one-dimensional Bose-Hubbard chains. We have obtained the phase diagram in the relevant parameter region where multiple phases are realized as a function of the intra- and inter-species interactions. Exploring the long-range decay of one- and two-body correlation functions we have been able to identify quantum droplets with atomic or pair superfluid phase coherence. We have determined a parameter region where three-body recombination effects are negligible, thus opening a way to produce long-lived bosonic droplets. Finally, we have found that the bipartite entanglement entropy present in the drops reaches a maximum when the central density saturates to the equilibrium one and then it decreases for increasing number of particles.

{\bf Acknowledgments.}
The authors thank Leticia Tarruell for useful comments and discussions and Joan Martorell for a careful reading of the manuscript. This work is partially funded by MINECO (Spain) Grants No. FIS2017-87534-P and FIS2017-84114-C2-1-P. 

\bibliographystyle{apsrev4-1}
\bibliography{paperbib}

\end{document}